\documentclass[10pt,twocolumn,letterpaper]{article}

\usepackage{cvpr}
\usepackage{times}
\usepackage{epsfig}
\usepackage{graphicx}
\graphicspath{ {./figures/} }
\usepackage{amsmath}
\usepackage{amssymb}
\usepackage{fancyhdr}
\fancypagestyle{empty}{%
  \fancyhf{} % clear all header and footers
  \cfoot{\thepage} % right align the page number in the footer
}

\usepackage[pagebackref=true,breaklinks=true,colorlinks,bookmarks=false]{hyperref}

\cvprfinalcopy % *** Uncomment this line for the final submission

\def\cvprPaperID{****} % *** Enter the CVPR Paper ID here

\ifcvprfinal\fi
\begin{document}
% \pagenumbering{⟨arabic⟩}

%%%%%%%%% TITLE
\title{Microvasculature Segmentation in Human BioMolecular Atlas Program (HuBMAP)}

\author{James Scanlon\\
{\tt\small jscanlon8@gatech.edu} \\
\and
Lisa D’lima\\
{\tt\small  ldlima3@gatech.edu} \\
\and
Yongqiang Wang\\
{\tt\small ywang4226@gatech.edu} \\
\and
Youssef Sultan\\
{\tt\small ysultan@gatech.edu}\\
% For a paper whose authors are all at the same institution,
% omit the following lines up until the closing ``}''.
% Additional authors and addresses can be added with ``\and'',
% just like the second author.
% To save space, use either the email address or home page, not both
% \and
% Second Author\\
% \and
% Georgia Institute of Technology\\
% North Avenue, Atlanta, GA 30332 US
% {\tt\small secondauthor@i2.org}
}

\maketitle
%\thispagestyle{empty}

%%%%%%%%% ABSTRACT
\begin{abstract}
   % The ABSTRACT is to be in fully-justified italicized text, at the top
   % of the left-hand column, below the author and affiliation
   % information. Use the word ``Abstract'' as the title, in 12-point
   % Times, boldface type, centered relative to the column, initially
   % capitalized. The abstract is to be in 10-point, single-spaced type.
   % Leave two blank lines after the Abstract, then begin the main text.
   % Look at previous CVPR abstracts to get a feel for style and length. 
   % The abstract section should contain a brief summary of your work that
   % includes the problem statement, proposed solution and results.
   Image segmentation serves as a critical tool across a range of applications, encompassing autonomous driving's pedestrian detection and pre-operative tumor delineation in the medical sector. Among these applications, we focus on the National Institutes of Health's (NIH) Human BioMolecular Atlas Program (HuBMAP), a significant initiative aimed at creating detailed cellular maps of the human body. In this study, we concentrate on segmenting various microvascular structures in human kidneys, utilizing 2D Periodic Acid-Schiff (PAS)-stained histology images. Our methodology begins with a foundational FastAI U-Net model, upon which we investigate alternative backbone architectures, delve into deeper models, and experiment with Feature Pyramid Networks. We rigorously evaluate these varied approaches by benchmarking their performance against our baseline U-Net model. This study thus offers a comprehensive exploration of cutting-edge segmentation techniques, providing valuable insights for future research in the field.
\end{abstract}

%%%%%%%%% BODY TEXT
\section{Introduction}
% \subsection{Background}
\subsection{Background \& Motivation}

The Human BioMolecular Atlas Program (HuBMAP) is an aspiring initiative funded by the National Institutes of Health (NIH). The program's goal is to develop an open, global framework that maps the human body at an unprecedented level of detail. HuBMAP lays the groundwork for breaking through medical enhancements in biology and disease starting with understanding the intricacies of segmentation. 

A key component for the HuBMAP initiative is the segmentation of high-resolution histology images. This task is particularly challenging because of how complex and ambiguous microvasculature is. Being able to segment microvasculature can give researchers a new perspective to analyze where the healthy cells are and 
how the cells influence a person’s health. 

In our paper, we focus on 2D PAS-stained histology slices of healthy human kidney tissues, of which only 1633 are annotated. Our aim is to initiate our study with an established U-Net baseline model. We will then advance to exploring and integrating more recent, innovative segmentation techniques. Through out experiments, we aim to improve the microvasculature segmentation in kidney histology images. Understanding vastly unknown cellular and microvasculature alterations can guide therapeutic strategies particularly in cancer or vascular diseases leading to more effective treatments and patient outcomes. 
%Explain Kaggle HubMAP competition background

%(5 points) What did you try to do? What problem did you try to solve? Articulate your objectives using absolutely no jargon. 
%Can also mention short description of dataset; will be covered with more depth in Segmentation Theory section

%(5 points) How is it done today, and what are the limits of current practice?

%(5 points) Who cares? If you are successful, what difference will it make? 

%-------------------------------------------------------------------------
%%%%%%%%% BODY TEXT
\subsection{Segmentation}
%(5 points) What data did you use? Provide details about your data, specifically choose the most important aspects of your data mentioned \href{https://arxiv.org/abs/1803.09010}{here}. You don’t have to choose all of them, just the most relevant.
The goal of segmentation is to identify boundaries and isolate distinct objects within an image. This is different from object detection which involves recognizing and classifying multiple objects of varying types within the same image. Segmentation identifies classes of pixels in a given image which also reflects the boundaries of different object groups. There are two types of segmentation, namely semantic and instance segmentation. Instance segmentation needs to identify all instances of a given class of objects, and semantic segmentation only needs to provide a single mask for a given class without distinguishing instances. The medical image segmentation usually falls into semantic segmentation where the interest lies in identifying certain organs from an image.

Deep learning has been applied to many fields including segmentation since its uprising from around 2015 \cite{deep_learning}. Numerous architectures have been proposed for segmentation tasks including FCN, UNet, DeepLab and Transformer based models. Fully convolutional networks (FCN) were first proposed by Long et al. (2015) \cite{fcn} which used a single upsampling convolution layer near the end to output pixel-wise segmentation masks. While, UNet was proposed to gradually upsample the feature maps with residual connections \cite{unet} improving upon the vanishing gradient problem during back propagation. This marked the explosion of encoder-decoder variants. The main theme of most variants is on adjusting the receptive field and residual connections. More residual connections during down/upsampling are proposed in V-net \cite{vnet}. Dense skip connections between encoder-decoder are added in UNet++ \cite{unet++, unet3+}. Atrous Spatial Pyramid Pooling (ASPP) and conditional random field (CRF) are proposed in DeepLabv3 to expand the receptive fields of convolutions and capture multi-scale contexts \cite{chen2017deeplab}. Then Aligned Xception and depthwise convolution are added in DeepLabv3+ \cite{chen2018deeplabv3} which enable feature map extractions at an arbitrary resolution. 

Transformers have gained popularity in sequential data processing like speech recognition and natural language processing \cite{vaswani2017attention, devlin2018bert, gpt}. They work by introducing an attention mechanism which learns global contextual information from inputs. Transformers have also been introduced into visual signal processing for the past few years with great success. The key idea is to divide images into patches which form a sequence and then apply the attention mechanisms as used in sequential data processing. A pure transformer architecture without convolutions is first proposed in \cite{vit} which achieves better accuracy than ResNet152 on various popular datasets. Shifted windows (Swin) is used to generate patches to provide information overlapping between patches \cite{liu2021swin}. The idea of Unet is combined with Swin to generate a U-shaped encoder-decoder architecture and similar architectures can be found in UNetR\cite{hatamizadeh2022unetr} and TransUnet \cite{chen2021transunet}. BEiT borrows the idea of masked tokens from the language model BERT \cite{devlin2018bert} and applies it to image transformer by tokenizing the image through autoencoding-style reconstruction \cite{bao2021beit} and masking out image patches.

\subsection{Dataset}
The dataset is from the Kaggle HuBMAP - Hacking the Human Vasculature: \href{https://www.kaggle.com/competitions/hubmap-hacking-the-human-vasculature/data}{kaggle.com/competitions/hubmap-hacking-the-human-vasculature}. It consists of 7033 colored TIFF images with resolution of 512x512 (of which 1633 are annotated for 3 classes) of human kidney histology slides being blood vessels, glomerulus, and unsure. The annotations stored in a json file represent coordinates of polygon masks that correspond to areas of each class on the image. The goal is to generate masks better known as pixel boundaries that can best predict areas of blood vessels within an image similar to how a subject expert would manually annotate such. 

Two of the most popular metrics for medical image segmentation are IoU and Dice score \cite{imgseg2021, wang2022medical}. IoU stands for intersection over union of the predicted masks and ground truth. Dice score is a similar metric which measures intersection over the summation of predicted masks and ground truth. For binary classes with the foreground as the positive class, the Dice score is identical to F1 score which measures precision and recall. Dice loss is usually used as loss function which is just $[1 - Dice \ Score]$.
\begin{equation}
    Dice(A, B) = \frac{2 \; | A \cap B|}{A + B} = \frac{2\; TP}{2\; TP + FP + FN} = F1
\end{equation}
%-------------------------------------------------------------------------
%%%%%%%%% BODY TEXT
\section{Medical Image Segmentation}
\subsection{Established U-Net Model}
For the purpose of medical imaging segmentation, UNet and DeepLab are the most pervasive models used due to their efficiency and ability to account for global and local features within medical images. In medical imaging, these models are able to account for minute details while also accounting for global context. While these models are both used in the segmentation, their architectures are very different: UNet implements a encoder-decoder structure that leverages skip connections, while DeepLab utilizes atrous convolutions that allow it to develop context for multiple levels of features. These atrous convolutions, while useful for capturing multiple levels of data, make DeepLab a more computationally expensive model than UNet to train, influencing our decision to use UNet within this project \cite{unetvsdeeplab}.  UNet's skip connections also help address the vanishing gradient problem, but are very computationally economical as they do not involve additional computation. With these two base models in mind, given that these models were published in 2015, there have been several iterations on them to account for improvements within the image segmentation space.
%https://ieeexplore.ieee.org/stamp/stamp.jsp?arnumber=9146648 UNET vs. DeepLab
\subsection{Recent Advances in Segmentation Techniques}
There are several models that make improvements on the UNet model, such as Multi-Scale UNet, which leverages multiple parallel branches to process the pixels at various levels of scope and obtain a more comprehensive segmentation while being more resilient to various sizes \cite{multiscaleUnet}.  This is different from UNet's original structure which only uses one branch within its model. Another improved version of the UNet model is Attention UNet, which adds self-attention gates to filter features propagated through the skip connections in the decoder \cite{attentionunet}. These self-attention gates learn a weighted sum  of all of the features in a given input, then dynamically determine their weights by importance. By leveraging attention gates, Attention UNet is able to  selectively hone in on the important features of an image and give them more weighting. Both Multi-Scale UNet and Attention UNet have been leveraged in medical imaging segmentation tasks and outperformed the existing UNet baseline model, showing that these models could prove useful in our task of segmenting human kidney tissues.
\section{Methodology and Experiments}
\begin{figure}[]
\begin{center}
\fbox{
   \includegraphics[width=0.9\linewidth]{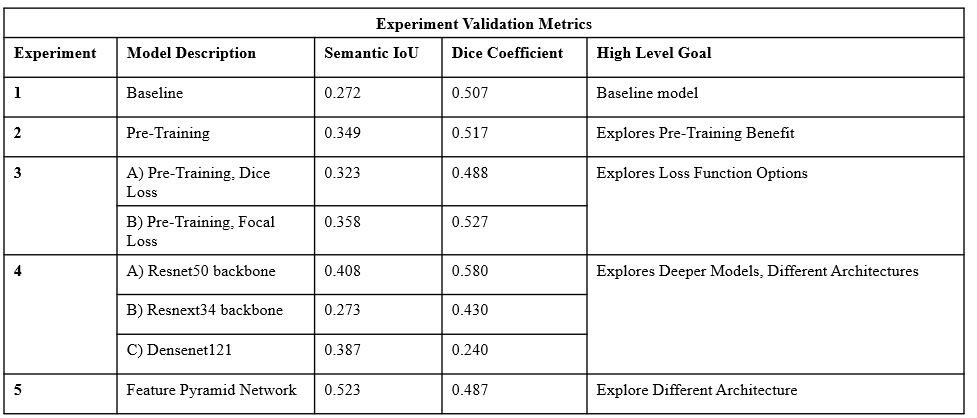}}
\end{center}
   \caption{Summary of Validation Metrics for Experiments}
\label{fig:summary}
\end{figure}

In our methodology, we initiated a series of five experiments each designed to uncover distinctive aspects of our deep learning model. From a baseline UNet, ResNet is added as encoder, focal loss is tested for imbalanced dataset, different networks are tested as encoders and finally FPN is introduced to improve the performance. All performance metrics are listed in Fig. \ref{fig:summary}.

\subsubsection{Experiment 1: Baseline UNet}
The initial experiment employed a baseline model, setting a reference point for subsequent comparison and evaluation. 

The annotations for the images, particularly those classified as 'blood vessels', are extracted and used to create a mask for each image. The mask initially is a zero-filled array which, upon validation of the relevant annotation, is filled accordingly. The mask is subsequently transformed into the required format using cv2 in order to align with the input dimensions of the model architecture. To ensure that only images with corresponding labels are included in our dataset, a filtration mechanism is incorporated. This ensures that our model only learns from images for which we have appropriate mask annotations and does not learn from annotated patterns that are not of interest.

The data loading and transformation process comprises image resizing, and normalization. Images are resized to a standard 128x128 format, and normalization is performed using the statistics from the ImageNet database.

The FastAI implementation of U-Net, which we've used as our baseline model, is built upon a modified version of the original U-Net architecture. One notable enhancement is the use of ResNet as the encoder, which provides the model with robust feature extraction capabilities due to ResNet's residual connections that mitigate the vanishing gradient problem during training. Specifically, in our baseline model, we employ ResNet34 as the encoder model of our U-Net. Another distinctive aspect of FastAI's U-Net is the integration of skip connections between the encoder and the decoder. These connections transfer feature maps from the encoder to the corresponding layers in the decoder, combining low-level detail-oriented features with high-level semantic features to yield more precise segmentation results. Finally, FastAI's U-Net model is highly flexible and can be easily adapted to exploring specific model optimizations. 

The model is initialized with a batch size of 8 and a learning rate of 1e-4, utilizing the Dice coefficient and Jaccard coefficient as performance metrics. An exploratory phase precedes the training phase, aimed at identifying the optimal learning rate for model training. Subsequently, the model is trained for 100 epochs.

Performance visualization follows the training phase, with loss values plotted to monitor the model's learning progression, identify potential overfitting or underfitting, and derive insights into possible areas of improvement. The performance of this baseline model provides a point of comparison against the various other architectures and methodologies explored in this study. The baseline achieves an IoU of 0.272 and Dice coefficient of 0.507.

% \begin{figure*}[t]
% % \begin{center}
% \fbox{
%    \includegraphics[width=0.9\linewidth]{figures/learning_curves_table.png}}
% % \end{center}
%    \caption{Experiment Learning Curves. From left to right: baseline model, model with focal loss to address class imbalance, model with pretrained Resnet50 as encoder, model with FPN.}
% \label{fig:lerning_curve}
% \end{figure*}

\begin{figure}[]
\begin{center}
\fbox{
   \includegraphics[width=0.9\linewidth]{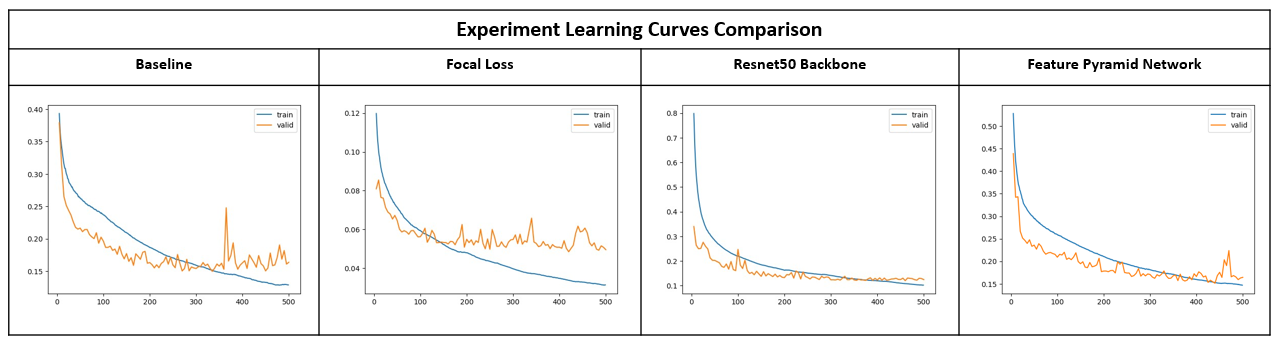}}
\end{center}
   \caption{Experiment Learning Curves. From left to right: baseline model, model with focal loss to address class imbalance, model with pretrained Resnet50 as encoder, model with FPN.}
\label{fig:lerning_curve}
\end{figure}

\subsubsection{Experiment 2: ResNet as encoder to UNet}
The second experiment in our series focused on the potential benefits of pre-training in deep learning models. In this approach, we leveraged the ResNet-34 model, initialized with weights obtained from prior training, to determine if such a strategy could speed up the training process and enhance the overall performance of the model. Our evaluation metrics, including Dice Loss and Intersection over Union (IoU), displayed notable improvement over our baseline model when applying this pre-training technique. This indicated a positive effect on the model's ability to segment the histology images accurately.

One of the main benefits of pre-training, particularly with relatively small healthcare datasets like ours, is the ability to mitigate the risk of overfitting. This often occurs when a model is excessively complex and begins to fit the idiosyncrasies of the training data rather than extracting generalizable features. Utilizing the ResNet-34 model, which was pre-trained on the vast ImageNet dataset, we leveraged a broad set of pre-existing features. These features range from simple constructs such as edges, found in the initial layers of the model, to more complex shapes in the deeper layers. Such a head-start facilitates a more effective and efficient learning process, enabling our model to converge faster and potentially achieve better performance than models trained from scratch. This improves the IoU from 0.272 to 0.349.

\subsubsection{Experiment 3: Loss Function for imbalanced class}
Our third experiment embarked on the task of comparing two distinct loss functions - Focal Loss and Dice Loss. Because the choice of loss function can significantly impact the model's learning process, this experiment was crucial to identifying that Focal Loss was the most suitable function for our specific task.  This experiment held particular significance due to the unique characteristics of our dataset and the specific task at hand. Our task involved the segmentation of irregularly shaped blood vessel regions within kidney histology images, a scenario compounded by a high class imbalance in-sample, and the presence of ambiguous, noisy boundaries around the blood vessels. 

One critical aspect of our dataset is the pronounced class imbalance, where the quantity of background data points vastly outnumbers the blood vessel regions of interest. The Focal Loss function is particularly designed for such scenarios. The model with focal loss yields a much lower loss, 0.06 vs 0.16 of baseline, as shown in Fig. \ref{fig:lerning_curve}. The model's focus on the "easy examples" are down-weighted whereas the "hard examples" are redirected to be focused on. Another layer of complexity is that our dataset has ambiguous and noisy boundaries around the blood vessels. Focal Loss tends to perform better than Dice Loss with noise, such as where the anatomical boundaries for segmentation are vague. The focal loss yields an IoU of 0.358 compared to 0.323 for Dice loss.

For future improvements, it may be beneficial to explore a customized loss function that integrates elements of Focal Loss and the Hausdorff distance\cite{burago2001course} and address overfitting. A customized loss function approach would penalize misclassified pixels in a more nuanced manner, taking into consideration their distance from the true boundaries of the anatomical regions. Adding regularization and data augmentation can mitigate overfitting from noise or outliers from the "hard examples" to classify.

\subsubsection{Experiment 4: Deeper Networks and Different Architectures}
In the fourth experiment, the investigation extended to both deeper and alternative architectures, diverging from the established baseline model that implemented a non-pretrained ResNet-34 structure. The architectures under consideration in this phase included ResNet-50\cite{he2015deep}, ResNeXt-34\cite{xie2017aggregated}, and DenseNet-121\cite{huang2018densely}. By increasing the complexity and depth of our model, we assessed the balance between model capacity and overfitting, aiming to optimize the model's performance.

Performance varied among the architectures. Specifically, ResNet-50 excelled, outperforming other Experiment 4 models in both IoU and Dice Coefficient. This could be attributed to the model's added complexity and depth compared to ResNet-34, enabling it to learn more intricate features that potentially enhanced segmentation quality and identification of ambiguous microstructures. Conversely, ResNeXt-34, was second place in Dice Coefficient, however registered the lowest score in IoU. Despite its architectural parallels with ResNets, ResNeXt-34’s grouped convolutional design encourages a more efficient exploration of the feature space but may not as effectively localize boundaries, hence impacting the IoU score. In the case of DenseNet-121, despite achieving a lower Dice Coefficient, it secured second place in terms of IoU. The direct connections between layers, a distinctive feature of DenseNet-121, aid in maintaining boundary details. This feature could enhance its IoU score, although potentially at the expense of a higher Dice score. The ResNet-50 yields the best IoU of 0.408 in this test.

\subsubsection{Experiment 5: Feature Pyramid Network}
In our concluding experiment, we implemented a custom Feature Pyramid Network (FPN) class that built upon FastAI UNet implementation. The FPN architecture was informed by the foundational work of Lin et al. \cite{seferbekov2018fpn} and exceeded the performance of all preceding models, including the unpretrained ResNet-34 U-Net baseline and the models assessed in Experiment 4. The key advantage of FPNs is their ability to capture features at different scales and merge them, creating a rich multi-scale feature representation. This is particularly beneficial in our application of segmenting kidney microvasculature where structures of interest can vary in size all within the same image.

\begin{figure*}[]
\begin{center}
\fbox{
   \includegraphics[width=0.9\linewidth]{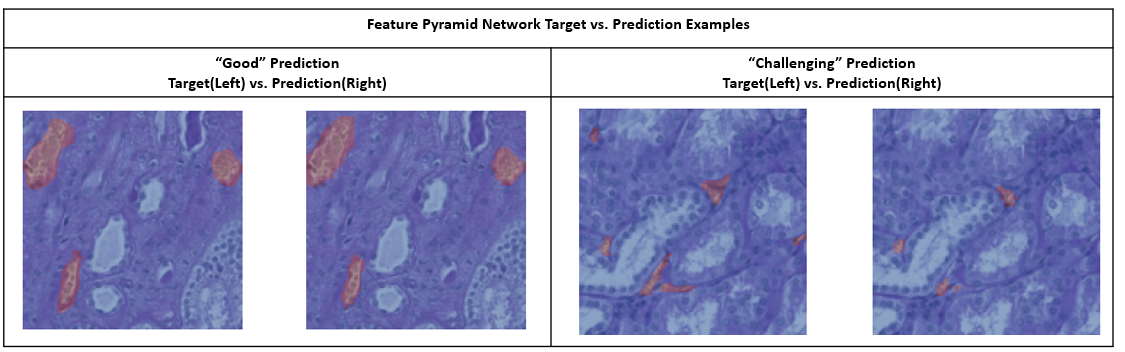}}
\end{center}
   \caption{Feature Pyramid Ground Truth Targets vs. Prediction outlining the "good" predictions(left) and "challenging" predictions(right) for significantly irregular blood vessels}
\label{fig:FP_resutls}
\end{figure*}

An FPN varies significantly from a conventional Convolutional Neural Network (CNN) in that it intentionally generates a multi-scale feature pyramid, utilizing feature maps from each pyramid level for the final task. This is achieved by incorporating a top-down pathway and lateral connections into the classic CNN architecture, which typically follows a bottom-up, feed-forward structure. The hierarchical, multi-scale nature of FPNs renders them exceptionally suited to our segmentation task. These networks are specifically designed to handle visual patterns across a range of scales and resolutions, thereby facilitating the segmentation of the complex, irregular blood vessels present in our high-resolution histology images. This architecture enables the model to capitalize on semantic features across all scales, thus enhancing performance for tasks such as object detection, where objects may appear at diverse scales within the image, and segmentation, where the accurate delineation of object boundaries often necessitates both high-level contextual understanding and low-level detail. The FPN is able to achieve the highest IoU of 0.523 among all tested models, yet still has challenges predicting significantly irregular blood vessels as showing in \ref{fig:FP_resutls}.

\begin{figure}[]
\begin{center}
\fbox{
   \includegraphics[width=0.9\linewidth]{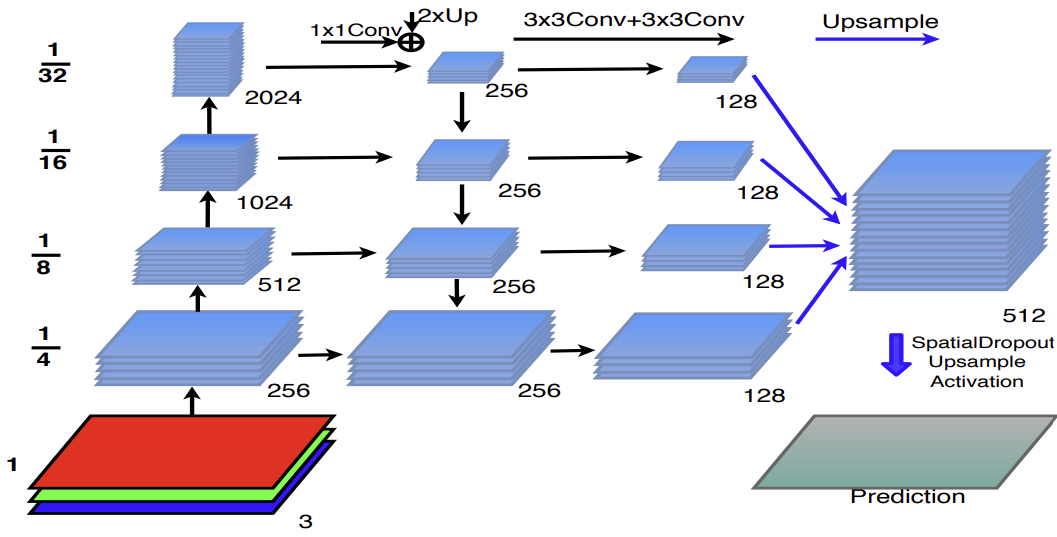}}
\end{center}
   \caption{Feature Pyramid Architecture from Figure 2 of \cite{seferbekov2018fpn}}
\label{fig:FPA}
\end{figure}

% \subsubsection{Experiment 4}

%Can add additional experiments as needed.

% \subsection{Results}
%Graphs
%(10 points) How did you measure success? What experiments were used? What were the results, both quantitative and qualitative? Did you succeed? Did you fail? Why? Justify your reasons with arguments supported by evidence and data.

%-------------------------------------------------------------------------

\section{Model Explorations}
\subsection{Standard UNet vs. FastAI Unet }
The original UNet \cite{unet} is implemented in PyTorch to explore potential architectural improvements, however the performance is not as good as expected. One possible cause is that since only a small portion of each image are the labeled class (blood vessels), the dataset is highly imbalanced. Initial test with standard cross entropy loss and Dice loss generated masks with almost all negatives, which means the model is only predicting false negatives. It was also shown in \cite{med_loss_func} that binary cross entropy loss and Dice loss do not seem to provide good results for imbalanced dataset. So the below weighted cross entropy loss is used to improve the performance.
\begin{equation}
    WCE(p, \hat{p}) = - \beta p \log \hat{p} - (1- \beta)(1-p) \log (1-\hat{p})
\end{equation}

\begin{figure}[]
\begin{center}
\fbox{
   \includegraphics[width=0.9\linewidth]{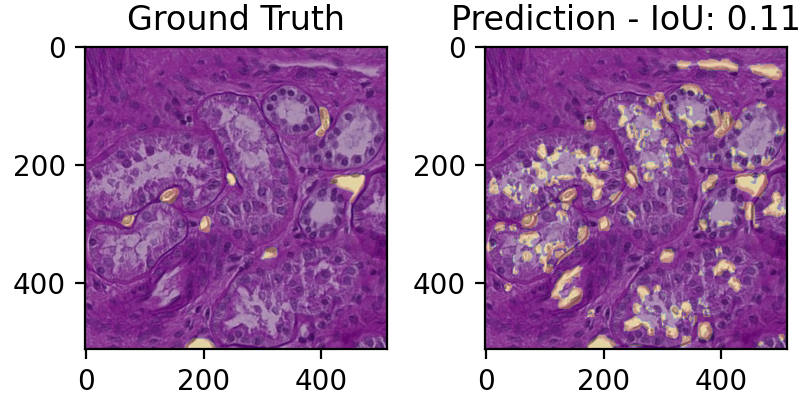}}
\end{center}
   \caption{Example of blood vessel segmentation with weighted binary cross entropy loss. The right shows the predicted mask which shows a lot of areas of false positives. }
\label{fig:wce_beta0.9}
\end{figure}

$\beta$ is set to 0.9 and the model stats to generate some meaningful predictions as shown in Fig. \ref{fig:wce_beta0.9}. This time there are more false positives with areas incorrectly labeled as blood vessels. The final IoU is only 0.11. The model seems to label a lot of areas with similar boundaries as blood vessels which indicates that wrong features could be learned by the model or boundaries are ignored and only inner patterns are learned. It would be more beneficial to enlarge the receptive field of convolutions to capture more information about the boundaries. It should be noted that the model uses batch size of 4 due to limitation of GPU memories, the batch normalization layers during convolutions were removed from the original UNet since small batch sizes tend to generate large variances in the loss gradients and make the training unstable. It should also be noted that fastai's UNet implementation uses a pretrained model as an encoder which generates feature maps as inputs to the UNet. This type of transfer learning could significantly improve the performance as proven here by comparison to our pure UNet implementation.

\subsection{Transformer based - SegFormer}
\begin{figure}[]
\begin{center}
 \fbox{
    \includegraphics[width=0.9\linewidth]{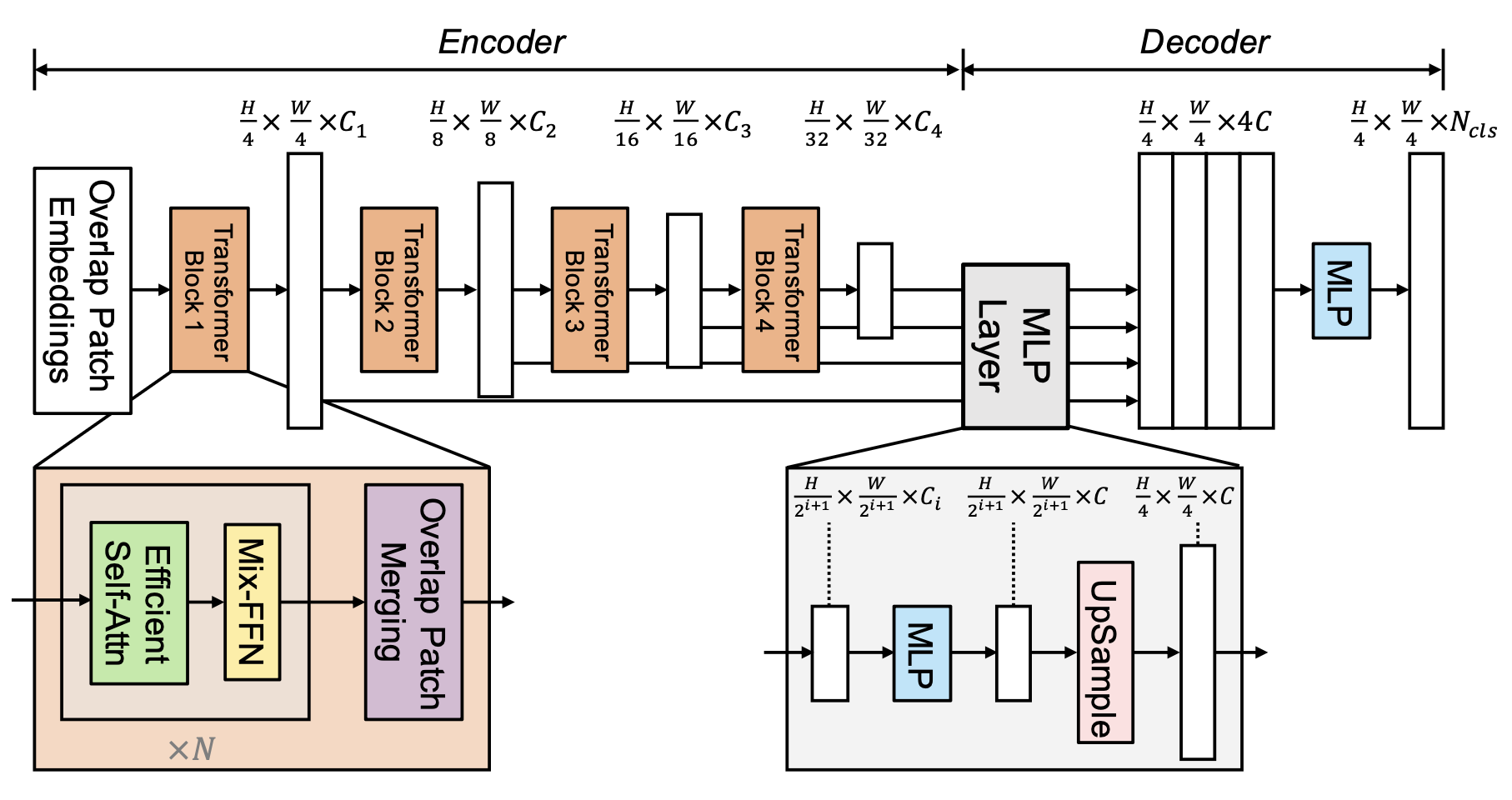}}
 \end{center}
    \caption{Architecture of a transformer based model for segmentation - SegFormer \cite{xie2021segformer}. }
\label{fig:segformer}
 \end{figure}

More traditional segmentation model architectures, such as U-Net, are primarily local operators that extract features based on local neighborhoods in an image. Even though there are many techniques like atrous convolutions used in DeepLab \cite{chen2018deeplabv3} to expand the receptive field, transformers offer a much better way to capture the global context of an image through the attention mechanism, regardless of their spatial distance. This could enable Transformers to capture more long range dependencies in the data in our kidney cell segmentation application. This motivates us to explore transformer based models for this dataset. SegFormer from \cite{xie2021segformer} is selected and the architecture is shown in Fig. \ref{fig:segformer}. It uses an attention mechanism which solves the problem with limited receptive fields in CNNs. Feature maps are extracted at each stage and concatenated by the decoder to output the final segmentation mask which is very similar to FPN \cite{seferbekov2018fpn}. Positional encoding is not needed because the author showed that zero padding can leak location information to the transformer. The weighted cross entropy loss is used. Due to limitation of computational resources, the network is downsized significantly from the original paper, including less number of stages and layers in the encoders, and higher downsamping/upsampling ratios. Because of this, the final IoU is still close to 0.1. However, we still believe transformer-based models have immense potential for many fields including segmentation. There could be a few reasons why it does not work very well for this dataset. One is that blood vessels masks in the training set show strong local patterns which transformers are less effective on compared to CNNs. Also due to limited time and resources, the model might be too small to extract all useful features.

% \subsection{Training}
% One of the challenges we dealt with was GPU Memory; the GPU we used to run our model could not manage more than 20GB of memory; given that we are dealing with large image files, this limited the batch size we could use for a given epoch.

% \section{Other Sections}

\section{Future Improvements}
There are several ways to improve tested models, with the first option being to use semi-supervised methods. The trained model can be used to generate pseudo-labels for them and feed back into training for data augmentation. Second, the size of failed models can be increased with more computational resources to improve the performance. Lastly, transfer learning has great potential as shown in the FastAI model which can also be introduced. 

\section{Conclusion}
Our study provided a comprehensive examination of various models and methodologies for segmenting microvasculature in 2D PAS-stained histology slices of healthy human kidney tissues. We began with a U-Net baseline model equipped with a ResNet-34 encoder, investigating the potential advantages of pre-training and the implementation of advanced loss functions such as Focal Loss. This was pertinent given the distinctive characteristics of our dataset, characterized by high class imbalance, irregularly shaped blood vessel regions, and ambiguous, noisy boundaries.

Exploring deeper and alternative architectures, namely ResNet-50, ResNext-34, and DenseNet-121, we were able to highlight the nuanced relationship between model depth, architecture, and performance on the given segmentation task. One particular note was the superior performance of the ResNet-50 model, emphasizing the potential benefits of deeper architectures. Our final experiment involved the integration of a Feature Pyramid Network (FPN) with a ResNet-34 encoder, which yielded the most effective results among all the methodologies we tested. The inherent multi-scale feature handling capacity of FPNs was particularly beneficial for our complex segmentation task, resulting in improved performance.

The task of segmenting high-resolution histology images presents considerable challenges, given its inherent variability and complexity. The superior performance of the FPN architecture in this context underscores the critical role of multi-scale feature integration for segmentation tasks and offers promising avenues for future research. Despite the promising results, several areas of potential improvement and further investigation exist. Future endeavors may include customization of loss functions, such as a hybrid approach combining Focal Loss with Hausdorff distance, examination of a wider variety of architectures such as Transformers, or fine-tuning the parameters of the FPN model. Our findings underline the significant potential of innovative deep learning methodologies in histological image segmentation, an essential component of the HuBMAP initiative. This research constitutes a key contribution towards the ultimate goal of HuBMAP: to create detailed cellular maps of the human body, a project with broad implications for future advancements in our understanding of biology and disease.

{\small
\bibliographystyle{ieee_fullname}
\bibliography{egpaper_for_review}
}

\end{document}